\documentclass[a4paper]{jpconf}
\usepackage{graphicx}
\usepackage{bm}
\begin{document}
\title{Effects of Lorentz boosts on Dirac bispinor entanglement}

\author{Victor A. S. V. Bittencourt}
\ead{vbittencourt@df.ufscar.br}
\address{~Departamento de F\'{\i}sica, Universidade Federal de S\~ao Carlos, PO Box 676, 13565-905, S\~ao Carlos, SP, Brasil.}
\author{Alex E. Bernardini}
\ead{alexeb@ufscar.br}
\address{~Departamento de F\'{\i}sica, Universidade Federal de S\~ao Carlos, PO Box 676, 13565-905, S\~ao Carlos, SP, Brasil.}
\author{Massimo Blasone}
\ead{blasone@sa.infn.it}
\address{Dipartimento di Fisica, Universit\`a degli studi di Salerno, Via Giovanni Paolo II, 132 84084 Fisciano, Italy}
\address{Also at: INFN Sezione di Napoli, Gruppo collegato di Salerno, Italy}

\begin{abstract}
In this paper we describe the transformation properties of quantum entanglement encoded in a pair of spin 1/2 particles described via Dirac bispinors. Due to the intrinsic parity-spin internal structure of the bispinors, the joint state is a four-qubit state exhibiting multipartite entanglement, and to compute global correlation properties we consider the averaged negativities over each possible bi-partition. We also consider specific bipartitions, such as the spin-spin and the particle-particle bipartitions. The particle-particle entanglement, between all degrees of freedom of one particle and all degrees of freedom of the other particle, is invariant under boosts if each particle has a definite momentum, although the spin-spin entanglement is degraded for high speed boosts. Correspondingly, the mean negativities are not invariant since the boost drives changes into correlations encoded in specific bipartitions. Finally, the results presented in the literature about spin-momentum entanglement are recovered by considering the projection of bispinorial states into positive intrinsic parity, and some striking differences between the appropriate approach for this case and the one usually treated in the literature are discussed.
\end{abstract}

\section{Introduction}



In the last decades many researches have been devoted to describe and contextualize quantum entanglement in relativistic setups \cite{relat01,relat02,relat03,relat04,relat05,relat06,relat07,relat08,relatvedral}. In particular, the implementation of information protocols, such as clock synchronization \cite{clock}, requires the description of the effects of frame transformations in quantum correlations, and since the beginning of the 2000’s many studies described how entanglement encoded in a pair of spins changes under Lorentz boosts  \cite{relat01, relat02, relat03, relat04, relat05, relat06, relat07, relat08, relatvedral}. The construction of spin states in this context, in its majority, follows the classification of the irreducible representations (irreps) of the Poincar\'{e} group \cite{fonda, wtung}, and the effect of a Lorentz boost is given by a momentum dependent rotation of the spin \cite{wigner}, which is the basis of some of the most iconic results in the field, such as the non-invariance of the reduced spin entropy of a single particle \cite{relat02}.

Although the setup usually adopted to describe transformation properties of quantum entanglement has given some interesting insights into the physics of relativistic quantum information, when massive charged fermions are considered as the physical carriers of spin 1/2, a more complete description of the problem is required. The physical particles, such as electrons and muons, are described by quantum electrodynamics, a theory which, apart from the usual Poincar\'{e} symmetry, also exhibit invariance under parity transformation \cite{wtung, weinberg}. This last symmetry operation exchange two irreps of the Poincar\'{e} group, and a proper description of such carriers of spin is given in terms of the irreps of the so called complete Lorentz group \cite{wtung}. The state of the particles are then described by four component objects, the Dirac bispinors, which satisfy the Dirac equation.

The solutions of Dirac equation were previously considered in the information theory framework to discuss the definition of spin operators in relativisitc quantum mechanics \cite{spins}, and the effects of Lorentz boosts on the entanglement encoded in superposition of Dirac equation solutions were considered for some specific states in connection with Wigner rotations \cite{bi-spinorarxiv02} and in discussing spin-spin entanglement in the context of the Fouldy-Wouthuysen spin operator \cite{bi-spinorFW}. However, due to the group structure intrinsic to the bispinors, such type of states carries two quantum bits \cite{SU2}, spin and intrinsic parity, and a given two particle state in this context is thus a four-qubit state.

In this paper we provide a general description of the changes driven by Lorentz boosts on quantum entanglement encoded in superpositions of two particle states described in terms of Dirac bispinors. We consider the effects of Lorentz boosts on the averaged entanglement encoded in each type of possible bi-partition of the system as well as in the entanglement shared among specific degrees of freedom (DoFs) of the system, for instance only between the spins or between the parities. In a first approach, we consider superpositions between spinor states where each particle has a definite momentum while the spins are superposed, the momenta are supposed to be (anti)parallel, and we consider both parallel and perpendicular boosts. For this scenario, we prove that the entanglement shared between all DoFs of one particle and all DoFs of the other particle is Lorentz invariant, despite the overall entanglement encoded in other types of bi-partitions is non-trivially affected by boosts. In a second scenario, we describe the effects of boosts on states with momentum superposition on a simplified framework and recover previous results of the literature about spin-momentum entanglement by considering projections into definite parities. We shown that boosts cannot create spin-momentum entanglement, a striking difference between our approach and previous ones based on the irreps of the Lorentz group. 

The paper is organized as follows. In Sec.~II the basic properties of Dirac bispinors, including intrinsic spin-parity entanglement, is reviewed. Section III introduces the two particle state without momentum superposition, and presents the different boost scenarios which will be considered. It is proved that the particle-particle entanglement is invariant under Lorentz Boosts although entanglement in other bi-partitions are not invariant. In Sec.~IV momentum superposition is introduced and the connection with results derived in the literature. Additionally, differences between the proper approach and the one adopted in the literature are discussed. To end up, Sec.~VI presents our last conclusion and future perspectives.

\section{Spin-parity entanglement }

Dirac equation was proposed as a wave equation invariant under Poincar\'{e} transformations and through which is possible to define a non-negative conserved probability current \cite{greiner}. In its Hamiltonian form, in the momentum space, Dirac equation reads
\begin{equation}
\label{diracequation}
H \, \psi = (\bm{p} \cdot \hat{\bm{\alpha}} + m \hat{\beta}) \psi = E_p \psi,
\end{equation}
where bold variables represent vector quantities with modulus denoted by $a = \vert \bm{a} \vert = \sqrt{\bm{a} \cdot \bm{a}}$, and $\hat{\alpha}_i$ and $\hat{\beta}$ (hats ``$~\hat{}~$'' denoting operators from hereafter) are anticommuting $n \times n$ matrices satisfying the relations
\begin{eqnarray}
\label{anticommuting}
\hat{\alpha}_i \hat{\alpha}_j + \hat{\alpha}_j \hat{\alpha}_i = 2 \delta_{ij} \hat{I}_4, &\mbox{ }& \hspace{0.5 cm} \hat{\alpha}_i  \hat{\beta} +  \hat{\beta} \hat{\alpha}_i =0, \nonumber \\
\hat{\beta}^2 &=& \hat{I},
\end{eqnarray}
with $\hat{I}$ is the identity operator. The matrices $\hat{\alpha}_i$ and $\hat{\beta}$ have different representations, interconnect via unitary transformations, and we shall adopt the Dirac representation in which
\begin{eqnarray}
\label{representation}
\hat{\alpha}_i = \left[ \begin{array}{cc} 0 & \hat{\sigma}_i \\ \hat{\sigma}_i &  0 \end{array}\right] \hspace{0.5 cm} \hat{\beta} = \left[ \begin{array}{cc} \hat{I} & 0 \\ 0 & - \hat{I} \end{array} \right].
\end{eqnarray}
The eigenstates of the Dirac Hamiltonian $\hat{H}= \bm{p} \cdot \hat{\bm{\alpha}} + m \hat{\beta}$, $u(p,s)$, with eigen-energy $+ E_p$, and $v(p,s)$ with eigen-energy $-E_p$, are the 4-component \textit{Dirac bispinors} given by
\begin{eqnarray}
\label{bispinors}
u (p, s) = \left[ \begin{array}{c} \sqrt{\frac{E_{p} + m}{2 E_{p}}} \chi_s  \\ \frac{\bm{p} \cdot \bm{\sigma}}{\sqrt{2 E_{p} (E_{p} + m)}} \chi_s \end{array}\right], \hspace{0.5 cm}  v (p, s) = \left[ \begin{array}{c} \frac{\bm{p} \cdot \bm{\sigma}}{\sqrt{2 E_{p} (E_{p} + m)}} \chi_s  \\    \sqrt{\frac{E_{p} + m}{2 E_{p}}} \chi_s \end{array}\right].
\end{eqnarray}
The two-component spinors $\chi_s$ ($s = \pm$) are related with the spin of the particle and their explicit form depends on the specific polarization adopted. In this paper we describe the bispinors as eigenstates of the Helicity operator $$\hat{h} = \frac{\bm{p} \cdot \hat{\bm{\Sigma}}}{\vert \bm{p} \vert},$$ with $$ \hat{\bm{\Sigma}} =  \left[ \begin{array}{cc} \hat{\bm{\sigma}} & 0 \\ 0 &  \hat{\bm{\sigma}} \end{array}\right]$$ the Pauli-Dirac spin operator. For such choice, $\vert \chi_s \rangle $ are the eigenstates of the operator $\bm{p} \cdot \hat{\bm{\sigma}}$ which, in terms of the eigenstates $\vert \pm \rangle$ of the $\hat{\sigma}_z$ operators, are given by
\begin{equation}
\label{helicityeigenstates}
\vert \chi_{\pm} \rangle = \frac{1}{\sqrt{2}} \left( \hat{I} \pm \frac{\bm{p}}{\vert \bm{p} \vert} \cdot \hat{\bm{\sigma}} \right) \vert \pm \rangle,
\end{equation}
and are orthonormalized $\chi_s^\dagger \chi_l = \delta_{s,l}$. The orthogonality relations satisfied by $u$ and $v$ are
\begin{eqnarray}
\label{orthogonality}
u^\dagger (p,s) u(p,r) =  v^\dagger (p,s) v(p,r) = \delta_{sr}, \hspace{0.5 cm} u^\dagger (p,s) v(p,r) = v^\dagger (p,s) u(p,r) =0,
\end{eqnarray}
and the completeness relation reads  $$\displaystyle \sum_{s}\Big[u(p,s) u^\dagger (p,s) - v(p,s)v^\dagger (p,s) \Big] = \frac{E_p + m}{E_p} \, \hat{I}.$$

The quantum information framework of Dirac equation can be set by noticing that in the representation (\ref{representation}), Dirac matrices are given in the form of tensor products, as to have $$\hat{\alpha}_i = \hat{\sigma}_x^{(P)} \otimes \hat{\sigma}_i^{(S)}, \hspace{0.5 cm} \hat{\beta} = \hat{\sigma}_z ^{(P)} \otimes \hat{I}^{(S)},$$ such that the Dirac Hamiltonian reads 
\begin{equation}
\label{twoqubithamiltonian}
H = \bm{p}\cdot (\hat{\sigma}_x^{(P)} \otimes \hat{\bm{\sigma}}^{(S)}) + m ( \hat{\sigma}_z ^{(P)} \otimes \hat{I}^{(S)}).
\end{equation}
Therefore, relativistic quantum mechanics can be interpreted as a two-qubit information theory associated with the discrete DoFs of the system: $P$, associated with the Hilbert space $\mathcal{H}_P$ describing the intrinsic parity DoF, and $S$, associated with the Hilbert space $\mathcal{H}_S$ describing the spin DoF \cite{SU2}. This two DoFs are related with the underlying group structure of the Dirac bispinors. The invariance under proper Lorentz transformations and space inversion, a symmetry present for example in quantum electrodynamics, requires the description in terms of the irreps of the complete Lorentz group \cite{wtung}, which contains, additionally to the spin, the intrinsic parity quantum number. In the case of spin $1/2$ particles (electrons, protons, muons, etc) such irreps belongs to $SU(2) \otimes SU(2)$ \cite{fonda, wtung}, as is explicitly shown in Eq. (\ref{twoqubithamiltonian}). In this context the bispinors (\ref{bispinors}) are rewritten as two-qubit states. For instance, the positive energy bispinor reads
\begin{equation}
\label{twoqubitspinor}
\vert \, u (p, s) \, \rangle =  \sqrt{\frac{E_{p} + m}{2 E_{p}}} \vert + \rangle_P \otimes \vert \chi_s \rangle_S +  \frac{1}{\sqrt{2 E_{p} (E_{p} + m)}} \vert - \rangle_P \otimes \,(\, \bm{p} \cdot \hat{\bm{\sigma}}^{(S)} \,  \vert \chi_s \rangle_S  \,),
\end{equation}
where $\vert \chi_s \rangle_S $ are the helicity eigenstates spinors (\ref{helicityeigenstates}), and we have introduced the subscripts P and S to indicate the intrinsic parity and spin spaces.

The generic form (\ref{twoqubitspinor}) is \textit{spin-parity} entangled. Entanglement is defined by means of the separability concept: A state with density matrix $\rho$ describing a system composed of two subsystems $A$ and $B$, each associated to its corresponding Hilbert space $\mathcal{H}_{A (B)}$, is separable if it can be written as \cite{reventanglement}
\begin{equation}
\rho = \displaystyle \sum_{i} c_i \, \rho_i^{(A)} \otimes \rho_i^{(B)},
\end{equation}
with $c_i \ge 0$, $\sum_i c_i = 1$, $\rho_i^{A(B)} \in \mathcal{H}_{A (B)}$. If $\rho$ is not a separable state, then it is entangled. Separability can also be defined in terms of the Peres criterion \cite{neg01}, which establishes that $\rho$ is separable iff the partial transpose density matrix $\rho^{A (B)}$, with respect to the $A$ subsystem, has only positive eigenvalues. With respect to a fixed basis on the composite Hilbert space $\{\vert \mu_i \rangle \otimes \vert \nu_j \rangle \}$ (with $\vert \mu_i \rangle \in \mathcal{H}_{A}$ and $\vert \nu_i \rangle \in \mathcal{H}_{B}$), the matrix elements of the partial transpose with respect to the first subsystem $\rho^{A}$ are given by
\begin{equation}
\langle \mu_i \vert \otimes \langle \nu_j \vert \rho^{A} \vert \mu_k \rangle \otimes \vert \nu_l \rangle = \langle \mu_k \vert \otimes \langle \nu_j \vert \, \rho \, \vert \mu_i \rangle \otimes \vert \nu_l \rangle.
\end{equation}
With Peres criterion one defines the bi-partite entanglement quantifier called Negativity $\mathcal{N}$ \cite{neg02}, given by
\begin{equation}
\label{negativity}
\mathcal{N}^{A; \, B}[\rho] =\frac{1}{d_i - 1}(\,  \displaystyle{\sum}_{i} \vert \mu_i \vert - 1 \, ),
\end{equation}
where $\mu_i$ are the eigenvalues of the matrix $\rho^{A}$, and $d_i = \mbox{dim}[ \, \mathcal{H}_A \,]$. Although many other entanglement quantifiers can be defined, negativity has the advantage to be computable without requiring any extremization process, and is the entanglement measurement adopted in this paper.

The intrinsic entanglement of Dirac bispinors was previously described for a plethora of scenarios \cite{diraclike01, diraclike02, aop01, aop02}. For example, a simple bidimensional scattering process of a plane wave by a step barrier can create entanglement between spin and parity DoFs in both reflected and transmitted waves \cite{aop01}. A more complete description of the influence of external fields on spin-parity entanglement was also studied \cite{aop02}, and its subsequent translation to Dirac-like systems was considered \cite{diraclike01, diraclike02}. For example, the tight binding model for bilayer graphene is formally equivalent to a modified Dirac Hamiltonian including external fields, and the single-particle excitations of the material were shown to exhibit a lattice-layer entanglement, analogous to the spin-parity entanglement of its corresponding relativistic bispinor description \cite{diraclike02}.

Different from the single particle spin-parity entanglement encoded in a single-particle, here we consider entanglement encoded among the DoFs of two bispinorial particles constructed in terms of tensor products of bispinors of the generic form
\begin{eqnarray}
\label{generalstate}
\frac{1}{\sqrt{N}} \displaystyle \sum_i^n c_i \, \vert \, u_A (p_i, \alpha_i) \rangle \otimes \vert \, u_B (q_i, \beta_i) \, \rangle,
\end{eqnarray}
where $N$ is a normalization factor. Due to the intrinsic structure of the bispinors (\ref{twoqubithamiltonian},\ref{twoqubitspinor}), the superposition (\ref{generalstate}) is a $4$-qubit state and although many effort has been devoted to devise how to proper characterize and quantify multipartite entanglement, up to now there is no precise method to accomplish such task \cite{reventanglement}. A simplified picture of the global measure of entanglement shared among different partitions of the system can be compute through averages over the possible bi-partitions \cite{rigolin}. For the 4-qubit states (\ref{generalstate}) there are $4$ types of bi-partitions: 
\begin{itemize}
\item $\{i; \, j, \, k, \, l \}$, for example the partition $\{S_1; \, P_1, \, P_2, \, S_2\}$, which is used to compute the entanglement between one the DoF and all the others;
\item $\{i,\, j; \, k, \, l \}$, which described the entanglement between pairs of DoFs. One type of such partition is $\{S_1, \, P_1; \, P_2, \, S_2\}$, which encodes the entanglement of all DoFs of one particle with all DoFs of the other;
\item $\{i; \, j, \, k\}$, obtained by tracing one of the DoFs. For example in the partition $\{S_1; \, P_2, \, S_2\}$ is encoded correlation between the spin of particle $1$ and all DoFs of particle $2$;
\item $\{i; \, j\}$, which encodes entanglement between only two DoFs. For example in the partition $\{P_1; \, P_2\}$ is encoded the entanglement between only the intrinsic parities.
\end{itemize}
In particular, to quantify the entanglement between spin of particle 1 and spin of particle 2 in a joint state $\varphi$, the partition $\{S_1; \, S_2\}$ should be considered, and entanglement would be given by $\mathcal{N}^{S_1; S_2}$, calculated with the reduced density matrix $\rho_{S_1, \, S_2} = \mbox{Tr}_{P_1, \, P_2}[\varphi \varphi^\dagger]$. Another possible approach to quantify the multipartite entanglement shared among the different DoFs of the system is to consider the geometry of the Hilbert space itself and compute different distance between the state and the set of so called K-separable states \cite{MassimoEnt}. Such procedure requires an involving extremization procedure and it will be postponed to future investigations.

As global measures of entanglement encoded in a state $\varphi$, one adopts the mean negativity in each bipartition as to have $4$ quantities defined as
\begin{eqnarray}
\label{meannegativities}
N^{(1)}&=& \bar{  \mathcal{N}}^{\, i; \, j\, k \, l}[\varphi]  \hspace{0.5 cm} N^{(2)} = \bar{  \mathcal{N}}^{\, i;\, j}[\varphi]  \nonumber \\
N^{(3)}&=& \bar{  \mathcal{N}}^{\, i; \, j\, k}[\varphi] \hspace{0.5 cm} N^{(4)}= \bar{  \mathcal{N}}^{\, i\, j; \, k\; l}[\varphi] .
\end{eqnarray}
For example, $N^{(1)} = (\mathcal{N}^{\, P_1; \, P_2\, S_1 \, S_2} + \mathcal{N}^{\, P_2; \, P_1\, S_1 \, S_2} + \mathcal{N}^{\, S_1; \, P_1\, P_2 \, S_2}+ \mathcal{N}^{\, S_2; \, P_1\, P_2 \, S_1})/4$. In fact, the relation between linear entropies and negativities for pure states sets an equivalence between $N^{(1)}$ and $N^{(4)}$, and the global measures of entanglement defined in \cite{rigolin}. 

\begin{figure}[h]
\centering
\includegraphics[width=  9 cm]{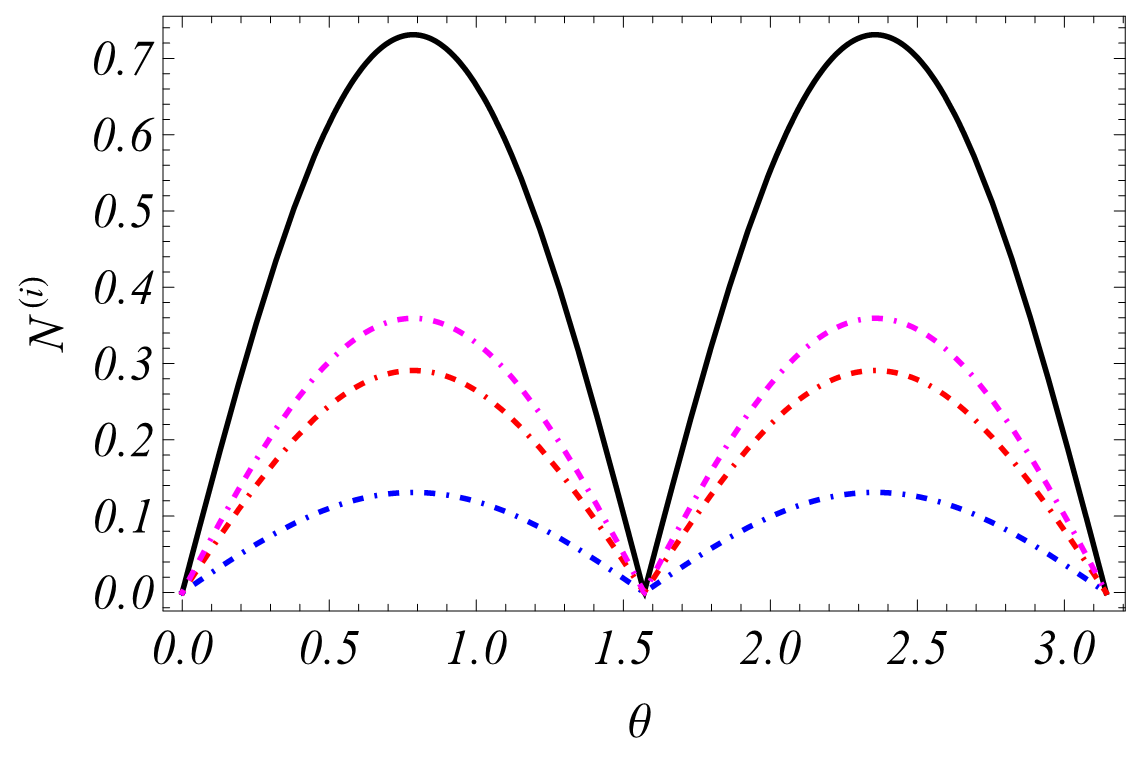}
\renewcommand{\baselinestretch}{1.0}
\caption{Mean negativities $N^{(i)}$ (\ref{meannegativities}) of the state (\ref{stategeneralized}) in the CoM frame $\bm{p}  = - \bm{q}$ as function of the superposition angle $\theta$. The initial rapidity of the particles is $\xi_0 = \mbox{arctanh}(E_p/ \vert \bm{p} \vert) = 1/2$ and the curves correspond to $N^{(1)}$ (black solid line), $N^{(2)}$ (red dashed line), $N^{(3)}$ (blue dot-dashed line) and $N^{(4)}$ (magenta dotted line). All quantifiers are maximed for maximal superpositions, for $\theta = \pi/4$ and $3 \pi/4$}
\label{F00}
\end{figure}

One of the simplest superpositions of the form (\ref{generalstate}) is the Bell-like state
\begin{equation}
\label{stategeneralized}
\Psi = \cos({\theta}) \, u_A(p, +) \otimes u_B(q, -) + \sin({\theta}) \,  u_A(p, -) \otimes u_B(q, +),
\end{equation}
where through this paper we describe the bispinors $u(p, s)$ polarized in the $e_z$ direction, such that $\vert \chi_{\pm} \rangle_S = \vert \pm \rangle_S$ in Eq. (\ref{twoqubitspinor}). The measures (\ref{meannegativities}) for the state $\Psi$ are depicted in Fig.~\ref{F00} for particles with opposite momentum and rapidity $\xi_0$ in the center of momentum (CoM) frame, i.e. $\bm{p} = - \bm{q} = ( 0,0, m  \sinh{(\xi_0)}) $, as function of the superposition parameter $\theta$. The entanglement measures are maximed for $\theta = \pi/4$, and $3 \pi/4$, which corresponds to a maximal superposition. 

\section{Lorentz boost of entangled bispinors}

Once the general framework is set, we now describe the effects of frame transformations on the entanglement content of a two bispinoral state. For a Lorentz transformation $\Lambda$ relating two frames $\mathcal{S}$ and $\mathcal{S}^\prime$ 
\begin{equation}
(\,x^\prime\,)^\mu = \Lambda ^\mu_{\, \nu}x^\nu,
\end{equation}
the covariant form of (\ref{diracequation}) is invariant if (for $\hat{\bm{\gamma}} = \hat{\beta} \hat{\bm{\alpha}}$ and $\hat{\gamma}^0 = \hat{\beta}$)
\begin{equation}
(\hat{\gamma}^\mu p_\mu -m \hat{I}) \psi(x) = 0 \rightarrow ((\hat{\gamma}^\prime)^\mu p^\prime_\mu -m \hat{I}) \psi^\prime(x^\prime) = 0,
\end{equation}
and thus the bispinor $\psi(x)$ transform as
\begin{equation}
\psi(x) \,  \rightarrow \, \psi^\prime(x^\prime) = \hat{S}[\, \Lambda \,] \psi(\Lambda^{-1} x^\prime),
\end{equation}
where $ \hat{S}[\, \Lambda \,]$ is the representation of the Lorentz transformation acting on the bispinor space.  For two inertial frames moving with respect to each other at a constant speed $\bm{v} = \vert \, \bm{v} \, \vert \, (e_x, e_y, e_z)$, the transformation $\Lambda$, a Lorentz boost, is given by
\begin{eqnarray}
[\Lambda (\omega)]_{ij} = \delta_{ij} + (\cosh{(\omega)} - 1)\,n_i \, n_j, \hspace{0.3 cm} [\Lambda (\omega)]_{i0} = [\Lambda(\omega)]_{0i} = \sinh{(\omega)} \,n_i,  \hspace{0.3 cm}  [\Lambda (\omega)]_{00} = \cosh{(\omega)}, \nonumber
\end{eqnarray}
where $\bm{n} = (n_x, \, n_y, \, n_z)$ is the unity vector specifying the direction of the boost and $\omega$ is the \textit{rapidity}
\begin{equation}
\omega = \mbox{arctanh}\left[ \, \frac{\vert \bm{v} \vert}{\sqrt{1 - \vert \bm{v} \vert^2}} \, \right].
\end{equation}

The representation of a boost in the bispinor space $\hat{S}[\, \Lambda(\omega) \,]$ is explicitly given by \cite{fonda, greiner}
\begin{equation}
\label{boostop}
\hat{S}[\Lambda(\omega)]= \exp \left[ - \frac{\omega}{2}\, \bm{n} \cdot \hat{\bm{\alpha}} \right] = \cosh{\left( \frac{\omega}{2} \right)} \hat{I} -\sinh{\left( \frac{\omega}{2} \right)} \bm{n} \cdot \hat{\bm{\alpha}},
\end{equation}
and bispinor $u$ therefore transforms as
\begin{eqnarray}
u(p,s) &\rightarrow&  \frac{1}{\sqrt{\cosh({\omega})}} \hat{S}[\Lambda(\omega)] u(p, s).
\end{eqnarray}
In particular, for two successive boosts with rapidities $\omega_1$ and $\omega_2$ in the same direction one has
\begin{equation}
 \hat{S}[\Lambda(\omega_2)] \hat{S}[\Lambda(\omega_1)] = \hat{S}[\Lambda(\omega_1 + \omega_2)],
\end{equation}
which can be explored to construct the general solutions (\ref{bispinors}) by performing a Lorentz boost from particles in the rest frame to the frame where the particle has momentum $\bm{p}$. In this case, the rapidity of the boost to construct the solutions from the rest frame is given by $\omega = \vert \bm{p} \vert / m$. For boosts parallel to the particles momentum, the transformation between $\mathcal{S}$ and $\mathcal{S}^\prime$ corresponds to just a change on the momentum of the state.

The action of $\hat{S}$ on superpositions of bispinors were previously considered mainly to discuss the definition of spin operators in relativistic quantum mechanics \cite{spins}. Due to the form of $\mathcal{S}$, some possible definitions of spin operators, apart from the usual Pauli-Dirac operator $1/2 \, \hat{\Sigma}$, other equally meaningful spin operators can be proposed. In particular, the Fouldy-Wouthuysen (FW) spin operator was considered to define a covariant spin reduced density matrix for Dirac bispinors as well as to define a proper position operator \cite{bi-spinorFW, celeri}. Additionally, considering superpositions of bispinors, as those of generic form (\ref{generalstate}), the effects of using different spin operators, with particular focus on eigenstates of the FW operator, were considered in the context of Bell's inequality \cite{bi-spinorFW}.

On the other hand, we shall consider the effects of boosts (\ref{boostop}) on the entanglement content of two bispinoral particles. In terms of two-qubit operators (\ref{boostop}) reads
\begin{equation}
\hat{S}[\Lambda(\omega)]=\cosh{\left( \frac{\omega}{2} \right)} \hat{I}^{(P)}\otimes \hat{I}^{(S)} - \sinh{\left( \frac{\omega}{2} \right)} \bm{n} \cdot( \, \hat{\sigma}_x^{(P)} \otimes \hat{\bm{\sigma}}^{(S)} \,),
\end{equation}
 and , from now on, we consider a simplified analysis in which, in the unboosted frame $\mathcal{S}$, the joint state is prepared as $\Psi$ (\ref{stategeneralized}) and address individually the effects of boosts parallel and perpendicular to the particle momenta. 

When a Lorentz transformation $\Lambda$ is performed, the state (\ref{stategeneralized}) transforms as
\begin{eqnarray}
\Psi \rightarrow \Psi^\prime &=& (\hat{S}[\Lambda] \otimes \hat{S}[\Lambda]) \Psi \nonumber \\
&=&  \cos({\theta}) \,\big (\, \hat{S}[\Lambda] u(p, +)  \, \big) \otimes \big (\, \hat{S}[\Lambda] u(q, -) \, \big) + \sin({\theta}) \, \big (\, \hat{S}[\Lambda] u(p, -) \, \big ) \otimes \big (\, \hat{S}[\Lambda]u(q, +) \, \big ), \nonumber
\end{eqnarray}
and the changing on entanglement between different partitions driven by the boost can be described by considering different partial transpositions and partial traces of the Lorentz transformed density matrix $\rho^\prime = \Psi^\prime ( \, \Psi^\prime \,)^\dagger$. In particular because the state is pure, negativities of the type $\mathcal{N}^{\, i; \, j\, k \, l}$ and $\mathcal{N}^{\, i \, j ;\, k \, l}$ shall exhibit the same behavior as the linear entropies $1- \mbox{Tr}_{i}[ \rho_i^2]$ and $1 -  \mbox{Tr}_{i, \, j}[ \rho_{i \, j}^2]$. For example, the entanglement in the bi-partition $\{P_1, \, S_1; \, P_2, \, S_2 \}$, which shall be called hereafter \textbf{particle-particle} entanglement, has the same behavior of the linear entropy $E_{L}=2( 1 - \mbox{Tr}[\rho^2_1])$, where
\begin{equation}
\rho_1 = \mbox{Tr}_{P_2, \, S_2} [\Psi \, \Psi^\dagger] =  \cos^2({\theta}) \, u(p, +) u^\dagger(p, +) + \sin^2({\theta})\,  u(p, -) u^\dagger(p, -),
\end{equation}
and through the orthogonality relations (\ref{orthogonality})
\begin{equation}
\label{linearentropyparticle1}
E_L=  \sin^2(\, 2 \theta\, ).
\end{equation}
Since
\begin{equation}
\label{neworth}
\big (\,\hat{S}[\Lambda]u(p, \alpha) \, \big )^\dagger \, \big( \, \hat{S}[\Lambda]u(p, \beta)\, \big) = \delta_{\alpha \, \beta},
\end{equation}
with respect to $\mathcal{S}^\prime$ one has the reduced density operator to the particle 1 given by
\begin{eqnarray}
\label{invariance}
\rho_1^\prime &=& \mbox{Tr}_{P_2, \, S_2} [\Psi^\prime \, (\, \Psi^\prime \, )^\dagger] \nonumber \\
 &=& \cos^2({\theta}) \, (\,\hat{S}[\Lambda]u(p, +)\,)\, (\,\hat{S}[\Lambda]u(p, +)\,)^\dagger + \sin^2({\theta})\,  (\,\hat{S}[\Lambda]u(p, -)\,) (\,\hat{S}[\Lambda]u(p, -)\,)^\dagger,
\end{eqnarray}
and, due to (\ref{neworth}), the same result (\ref{linearentropyparticle1}) holds. Therefore entanglement between all DoFs of particle $1$ and all DoFs of particle $2$ is invariant under Lorentz boosts. We point here that, although a specific state was considered, the invariance property derived here holds for states with no momentum superposition, i.e. the bispinors on the superposition associated with a given particle have all the same momentum, which have the general form

\begin{eqnarray}
\frac{1}{\sqrt{N}} \displaystyle \sum_{\alpha = \pm, \beta = \pm} c_i \, u(p, \alpha)  \otimes u(q, \beta).
\end{eqnarray}

Despite the invariance of particle-particle entanglement, other bi-partitions may have non-invariant entanglement, as is now addressed for boosts parallel and perpendicular to the $\bm{e}_z$ direction.

\subsection{Parallel boost}

\begin{figure}[h]
\centering
\includegraphics[width= 8 cm]{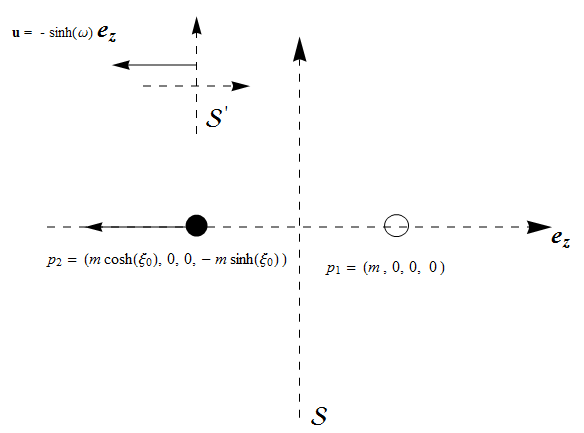}
\renewcommand{\baselinestretch}{1.0}
\caption{In the parallel Boost framework $\mathcal{S}$ corresponds to the rest frame of particle $1$, and in which particle $2$ moves with rapidity $\xi_0$ in the direction $-e_z$. The frame $\mathcal{S}^\prime$ moves with respect to $\mathcal{S}$ with rapidity $\omega$ in the $-e_z$ direction.}
\label{S02}
\end{figure}

The simplest boost framework can be constructed by considering the reference frame $\mathcal{S}$ as the rest frame of particle $1$ while particle $2$ is moving with rapidity $\xi_0 = \mbox{arcsinh}\, (\vert \bm{q} \vert/m)$ in the $-e_z$ direction. The 4-momenta $p$ and $q$ of particle 1 and 2 with respect to $\mathcal{S}$ are given by:
\begin{equation}
p = (m,0,0,0) \hspace{0.5 cm} q = (m \cosh{(- \xi_0)},0,0, m \sinh{(- \xi_0)}).
\end{equation}
The frame $\mathcal{S}^\prime$ in this setup moves with rapidity $\omega$ in the direction $- e_z$, such that the momenta with respect to $\mathcal{S}$ reads
\begin{equation}
p^\prime =  (m \cosh{(\omega)},0,0, m \sinh{(\omega)}) \hspace{0.5 cm} q^\prime = (m \cosh{(\omega - \xi_0)},0,0, m  \sinh{(\omega - \xi_0)}  ),
\end{equation}
and one notices that for a boost $\omega = \xi_0$, $\mathcal{S}$ is the rest frame of particle 2, while for a boost $\omega = \xi_0/2$, $\mathcal{S}^\prime$ is the CoM of the system. Additionaly, if $\xi_0 = 0$ then the particles are in rest with respect to each other. Figure~\ref{S02} depicts pictorically the framework described.

\begin{figure}[h]
\centering
\includegraphics[width=  15 cm]{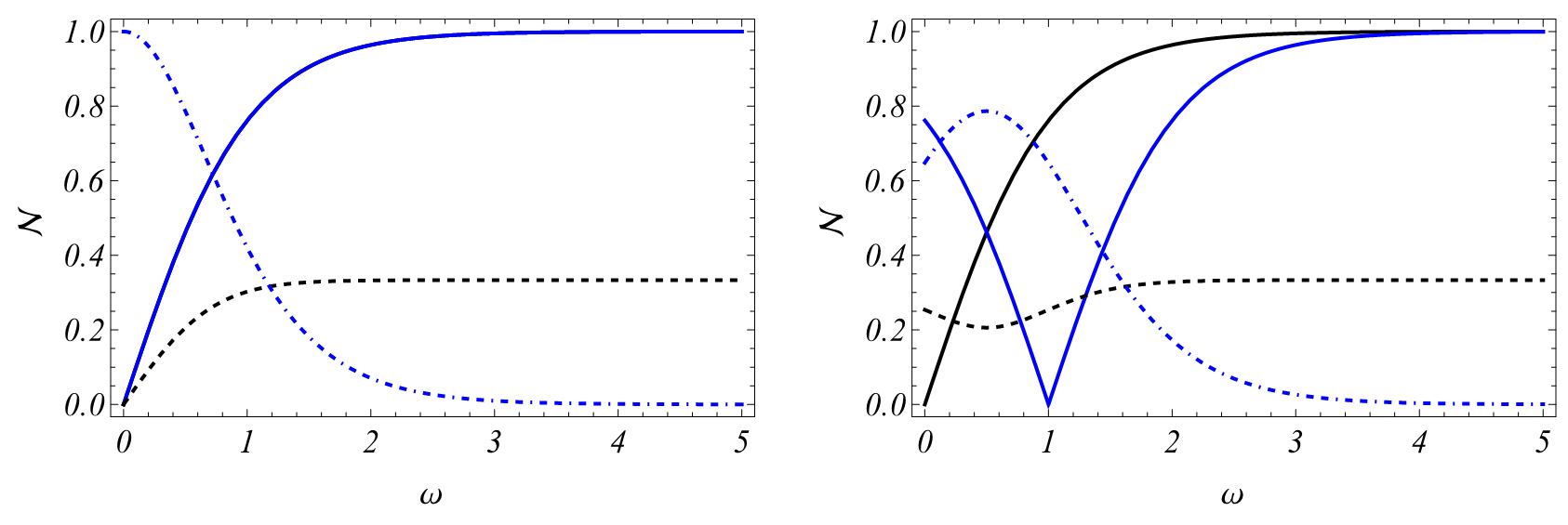}
\renewcommand{\baselinestretch}{1.0}
\caption{Negativities (\ref{negset01}) - $\mathcal{N}^{P_1; P_2 \, S_1 \,S_2}$ (black solid line), $\mathcal{N}^{P_2; P_1 \, S_1 \,S_2}$ (blue solid line), $\mathcal{N}^{S_1; S_2}$ (blue dot-dashed line) and $\mathcal{N}^{P_1\, P_2 ; S_1 \, S_2}$ (black dashed line) as function of the boosts rapidity $\omega$, for $\xi_0 = 0$ (left plot) and $\xi_0 = 1$ (right plot), and for $\theta = \pi/4$. In the situation where both particles are at rest with respect to each other, entanglement exhibit a monotonous behavior. While entanglement between parities and the rest of subsystems, and between both parities and both spins increases with the rapidity, the spins DoFs disentangle for high speed boosts. A similar behavior is shown when one of the particles moves with respect to the other in $\mathcal{S}$, although in this case entanglement is not a monotonous function of the rapidity. In particular, for $\omega= \xi_0$, parity $2$ disentangles for the rest of the system as for such boost $\mathcal{S}^\prime$ is the rest frame of particle $2$. Additionally, spin-spin entanglement is maximum at the CoM $\omega = \xi_0/2$, for which the entanglement between both parities and both spins reaches its minimum.}
\label{Neg01}
\end{figure}

The boost $\hat{S}[\Lambda]$ relating $\mathcal{S}$ to $\mathcal{S}^\prime$ is in the same direction of $\bm{q}$, and therefore it only changes the momentum of the corresponding bispinor, i.e.
\begin{equation}
u^\prime (q^\prime, s) = u (q^\prime, s).
\end{equation}
In this framework, one can explicitly compute the following negativities
\begin{eqnarray}
\label{negset01}
\mathcal{N}^{S_1; P_1, \, P_2, S_2}[\Psi^\prime] &=& \mathcal{N}^{S_2; P_1, \, P_2, S_1}[\Psi^\prime]= \vert \sin{2 \theta} \vert,  \nonumber \\
\mathcal{N}^{P_1 \, P_2; S_1 \, S_2}[\Psi^\prime] &=& \sqrt{1 - \mbox{sech}^2\,{\omega}\, \mbox{sech}^2\,{(\, \omega - \xi_0 \, )} } \, \frac{\vert \sin{2 \theta} \vert}{3}, \hspace{0.5 cm} \mathcal{N}^{P_1; P_2 \, S_1 \, S_2}[\Psi^\prime] =  \vert \, \tanh{\omega} \, \sin{2 \theta} \, \vert, \nonumber \\
\mathcal{N}^{S_1; S_2}[\Psi^\prime] &=& \,  \mbox{sech}\,{\omega}\, \mbox{sech}\,{(\, \omega - \xi_0 \, )} \, \vert \sin{2 \theta} \vert, \hspace{0.67 cm} \mathcal{N}^{P_2; P_1 \, S_1 \, S_2}[\Psi^\prime] =  \vert \, \tanh{(\, \omega - \xi_0 \,)} \, \sin{2 \theta} \, \vert, \nonumber \\
\end{eqnarray}
The partitions $\{ S_1; P_1, \, P_2, S_2 \}$ and $\{ S_2; P_1, \, P_2, \, S_1 \}$ exhibit invariant entanglement. Other bi-partitions are non-invariant as shown in Fig.~\ref{Neg01}, which depicts the negativities (\ref{negset01}) for a maximal superposition $\theta = \pi/4$ as function of the boost rapidity $\omega$ for $\xi_0 = 0$ (left plot) and $\xi_0 = 1$ (right plot). While entanglement between parity $1$ ($2$) and the rest of the subsystems increases with the boost rapidity, spin-spin entanglement vanishes in the limit $\omega \rightarrow \infty$, being a non-monotonous function of $\omega$ if the particles are not in rest with respect to each other, and exhibiting a maximum in the CoM. Additionally $\mathcal{N}^{P_2; P_1 \, S_1 \, S_2}$ exhibit a zero at $\omega = \xi_0$, which is expected since this point corresponds to the rest frame of particle 2. In this framework parity-parity entanglement vanishes.

\begin{figure}[h]
\centering
\includegraphics[width=  15 cm]{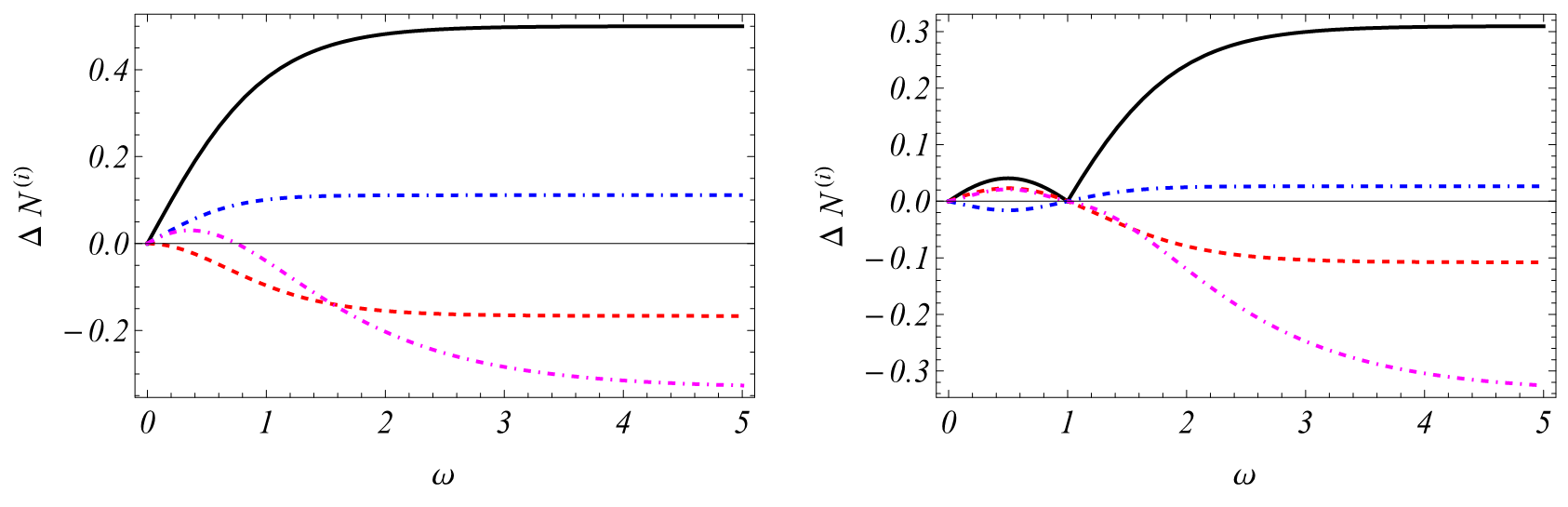}
\renewcommand{\baselinestretch}{1.0}
\caption{Variation of mean Negativities $\Delta N^{(i)} = N^{(i)}[\Psi^\prime] - N^{(i)}[\Psi]$ as function of the boost rapidity $\omega$, for $\xi_0 = 0$ (left plot) and $\xi_0 = 1$ (right plot), and for $\theta = \pi/4$. The curves correspond to $\Delta N^{(1)}$ (black solid line), $N^{(2)}$ (red dashed line), $N^{(3)}$ (blue dot-dashed line) and $N^{(4)}$ (magenta dotted line). One notice that all differences vanishes for $\omega= \xi_0$, when a Boost to the reference frame of particle $2$ is considered. Moreover, the average $N^{(4)}$ exhibit an invariance point even if both particles are at rest with respect to each other, and $ N^{(1)}$ always increases with the boost.}
\label{N00}
\end{figure}

The other partitions also have non-invariant negativities. For instance the negativity $\mathcal{N}^{P_i; S_1 \, S_2}$, quantifying the entanglement between parity $i$ and both spins is a non-monotonous function of the boost rapidity vanishing for $\omega \rightarrow \infty$ and in the frame where the $i$-th particle is at rest. The mean partition entanglements are thus not invariant, as shown if Fig.~\ref{N00} which depicts the variations of the mean negativites defined in (\ref{meannegativities}), $\Delta N^{(i)} = N^{(i)}[\Psi^\prime] - N^{(i)}[\Psi]$, for the different types of bi-partitions and for the same set of parameters of Fig.~\ref{Neg01}. The mean negativity over the partitions of the type $\{i; \, j, \, k, \, l\}$ always increases under boosts. For boosts with $\omega < \xi_0$, the mean entanglement in partitions of the type  $\{i; \, j, \, k, \, l\}$ and$\{i; \,j\}$ while the mean entanglement in the partition $\{i; \, j, \; k \}$ decreases. When the boost has a rapidity bigger than the rapidity of particle 2, one observe an increasing on the entanglement of partition $\{i; \, j, \, k, \, l\}$ and $\{i; \, j, \; k \}$, while the other two partitions exhibit a degradation in the mean entanglement. Among all partitions, $\{i; \, j, \; k\}$ has the smaller entanglement variation due to the boost.

\subsection{Perpendicular boost}

\begin{figure}[h]
\centering
\includegraphics[width= 8 cm]{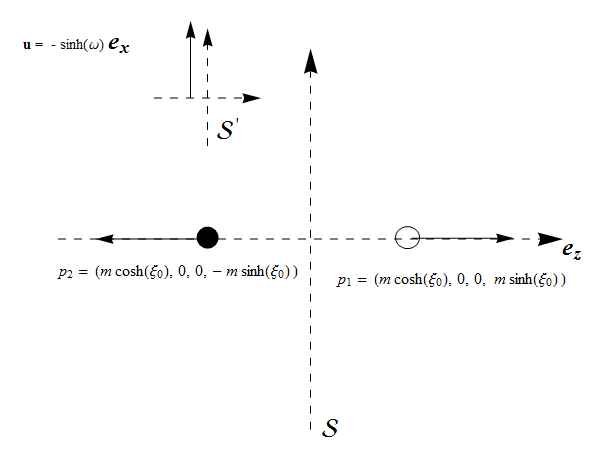}
\renewcommand{\baselinestretch}{1.0}
\caption{Perpendicular Boost framework. From the CoM frame $\mathcal{S}$, one performs a Lorentz boost with rapidity $\omega$ in the direction $e_x$, perpendicular to the particles momenta. Different from the previous scenario, the boost not only changes the momentum of the bispinors but also mixed its components, changing the entanglement between the different DoFs of the system.}
\label{S03}
\end{figure}

When a boost is performed in a direction different from the particles momentum, the components of the bispinor get mixed. In the perpendicular boost framework, $\mathcal{S}$ corresponds to the CoM in which the 4-momenta $p$ and $q$ reads
\begin{equation}
p = (m \cosh{(\xi_0)},0,0,m \sinh{(\xi_0)}) \hspace{0.5 cm} q = (m \cosh{(- \xi_0)},0,0, m \sinh{(- \xi_0)}).
\end{equation}
The inertial frame $\mathcal{S}^\prime$ moves in the $e_x$ direction with rapidity $\omega$ with respect to $\mathcal{S}$ and the transformed 4-momenta are given by
\begin{eqnarray}
p^\prime &=& (m \cosh{(\xi_0)} \cosh{(\omega)},m \cosh{(\xi_0)} \sinh{(\omega)},0,m \sinh{(\xi_0)}) \nonumber \\
q^\prime &=& (m \cosh{(- \xi_0)}  \cosh{(\omega)},m \cosh{(-\xi_0)} \sinh{(\omega)},0, m \sinh{(- \xi_0)}).
\end{eqnarray}
The bispinors transforms as 
\begin{eqnarray}
u(p, \pm) \rightarrow u^\prime &=&  \frac{1}{\sqrt{\cosh{\omega}}}\left[\, \cosh{\left( \, \frac{\omega}{2}\, \right)} \hat{I} - \sinh{\left( \, \frac{\omega}{2}\, \right) } \hat{\alpha}_x \right] u(p, \pm) \nonumber \\
&=& \frac{1}{\sqrt{\cosh{\omega}}}\left[ \begin{array}{c} \cosh{\left( \, \frac{\omega}{2}\, \right)} f(p) \chi_\pm \mp  \sinh{\left( \left( \, \frac{\omega}{2}\, \right) \right)} g(p) \chi_{\mp}  \\ \pm  \cosh{\left( \, \frac{\omega}{2}\, \right)} g(p) \chi_\pm - \sinh{\left( \, \frac{\omega}{2}\, \right) } f(p) \chi_\mp \end{array}\right],
\end{eqnarray}
where the shorthand notation $f(p) = \sqrt{\frac{E_p + m}{2 E_p}}$, $g(p) = \frac{p}{\sqrt{2 E_p (E_p + m)}}$ was adopted. Figure~\ref{S03} depicts schematically the Boost scenario.

\begin{figure}[h]
\centering
\includegraphics[width=  9 cm]{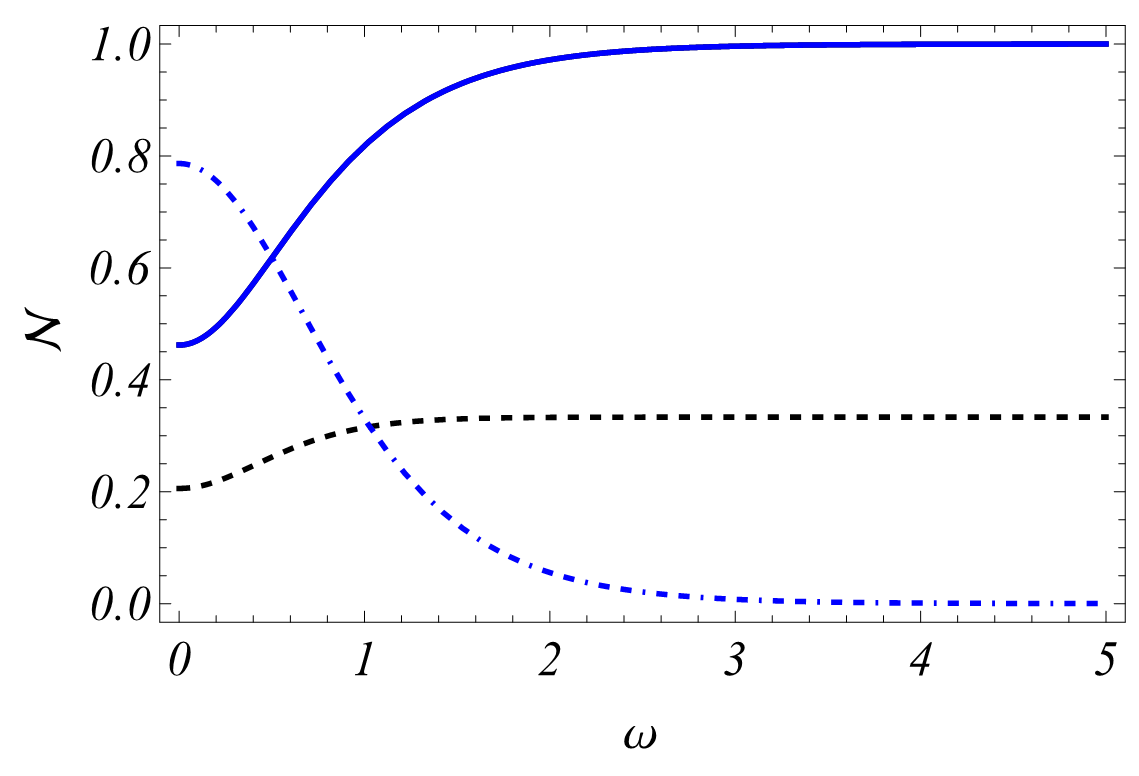}
\renewcommand{\baselinestretch}{1.0}
\caption{Negativities as function of the boost rapidity $\omega$ in the direction perpendicular to the momenta of the particles. The plot-styles are in correspondence with Fig.~\ref{Neg01} and $\mathcal{S}$ corresponds to the CoM of the particles, in which $\xi_0 = 1/2$. Entanglement exhibit a monotonous behavior as function of $\omega$, and while $\mathcal{N}^{P_{1 (2)}; \, P_{2 (1)}, \, S_1, \, S_2}$ and $\mathcal{N}^{P_{1}, \, P_{2}; \, S_1, \, S_2}$ increase with the boost rapidity, spin-spin entanglement is degraded. In the limit of high speed boost, $S_1$ and $S_2$ are separable, while the entanglement between parities and the other subsystems is maximum.}
\label{Neg03}
\end{figure}

Figure~\ref{Neg03} shows the same negativities depicted in Fig.~\ref{Neg01} but for the boost from the CoM in the $e_x$ direction . The qualitative behavior of the functions are the same as those depicted in the right plot of Fig.~\ref{Neg01}. Similar to the parallel boost from the common rest frame, the entanglement between spins is degraded by the boost and for boosts with high rapidity the spins of the particles are completely separable. On the other hand, the entanglement between one intrinsic parity and the other DoFs and between both parities and both spins are increasing functions of the boost rapidity, suggesting that the boost distribute the spin-spin entanglement among other partitions. Additionally, the correlations between one spin and the other DoFs are invariant under such boosts.

\begin{figure}[h]
\centering
\includegraphics[width=  9 cm]{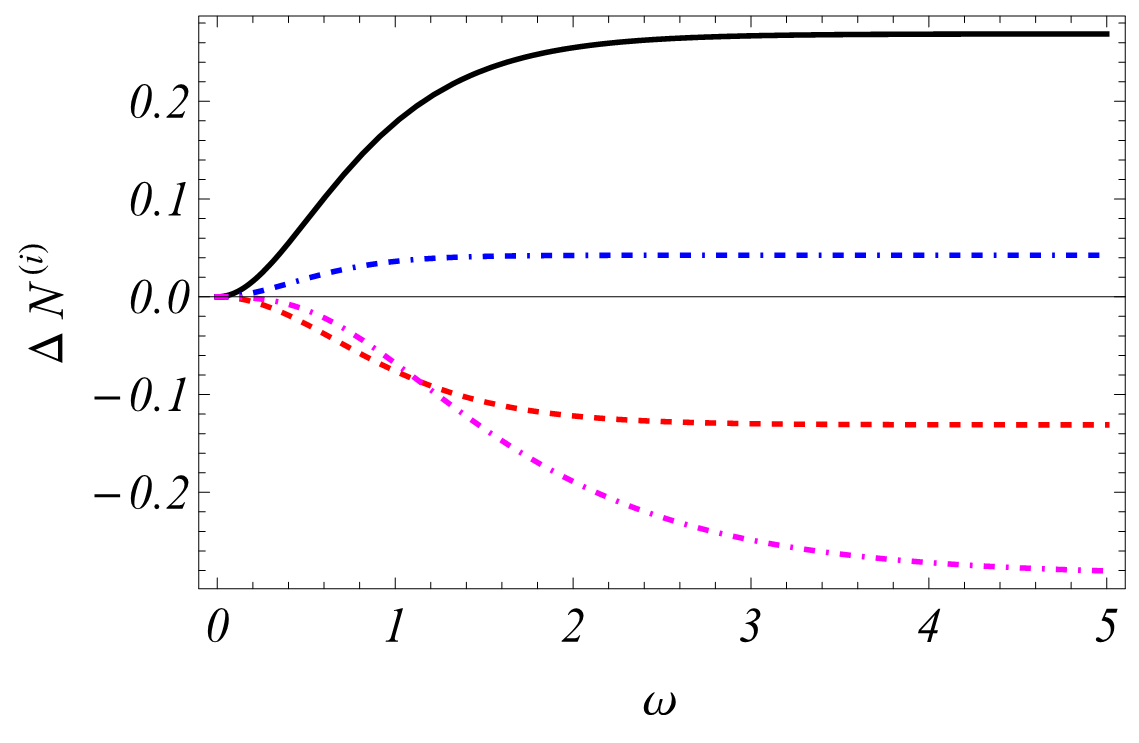}
\renewcommand{\baselinestretch}{1.0}
\caption{Variations of mean negativities $N^{(i)}$ as function of the boost rapidity $\omega$ for the perpendicular boost. The plot styles are in correspondence with those of Fig.~\ref{S03}. The variations on entanglement behave similar to those exhibited on parallel boost framework (see the left plot of Fig.~\ref{S03}). While the mean entanglement on the partitions $\{i; \, k, \, k, \, l\}$ and $\{i, \, j, \, k\}$ increase, the other two partitions have the entanglement degraded by the boost.}
\label{AveragesW}
\end{figure}

Different from the parallel boost, the mean negativities exhibit a monotonous behavior as function of the rapidity for a perpendicular boost, as depicted in Fig.~\ref{AveragesW} that shows the variations of the mean negativities $\Delta N^{(i)}$ as function of $\omega$ for $\theta = \pi/4$ considering that $\mathcal{S}_0$ is the CoM frame for which is supposed that $\xi_0 = 1/2$. The partitions $\{i; \, j, \, k, \, l\}$ and $\{i; \, j, \, k\}$ have increasing mean entanglement, while the other two partitions lost entanglement due to the boost. The behaviors exhibit in Figs.~\ref{N00} and \ref{AveragesW} suggest that a boost has an overall effect of increasing the mean entanglement in the partition $\{i; \, j, \, k, \, l \}$, that can be used as a global measure of entanglement for such pure states \cite{rigolin}. This general behavior is also observed for anti-symmetric states and more general boosts \cite{NewPRA}. A part from this general behavior, there is no compensation between degradation and increasing of the mean partition entanglement considered.

\section{Momentum superposition - Recovering the results in the Literature}

The behavior of entanglement under Lorentz boost was also described for spin states in momentum superposition \cite{ relat02, relat06, relat07}. In the framework of the irreps of the Lorentz group, the simplest separable two particle state in a momentum superposition is given by \cite{relat07}
\begin{equation}
\label{previous}
\vert \phi \rangle = \big(\,\cos({\alpha})\, \vert \bm{p} \rangle_1 \, \otimes \vert \bm{q} \rangle_2 + \sin({\alpha})\, \vert \bm{q} \rangle_1 \, \otimes \vert \bm{p} \rangle_2\, \big) \otimes \vert \phi_{spin} \rangle,
\end{equation}
where $ \vert \phi_{spin} \rangle$ is a joint spin state, and $\vert \bm{p} \rangle_i$ is the momentum state of the particle $i$. The state (\ref{previous}) is separable between spin and the momenta, and it was shown that a Lorentz boost entangles the spins and the momenta DoFs \cite{relat02, relat07} depending on the value of the Wigner rotation angle $\delta$, which for boosts from the CoM and perpendicular to the particle momenta is given in terms of the initial rapidity and the boost rapidity by $$\tan (\delta) = \frac{\sinh{(\xi_0)} \, \sinh{(\omega)}}{\cosh{(\xi_0)} + \cosh{(\omega)}}.$$ 

The momentum eigenstates $\vert \bm{p} \rangle$ have normalization given by
\begin{equation}
\int \frac{d^3 p}{2 E_p} \langle \bm{p} \vert \bm{q} \rangle = 1,
\end{equation}
and, if one considers a particle with highly concentrate momentum distribution, for instance if the state $\vert \bm{p} \rangle$ is associated with a gaussian distribution centered around $\bm{p}$ with a very small width, the distributions of the opposite momenta $\bm{p} = - \bm{q}$ will have a negligible overlap. With this assumption,  the momentum DoF introduced in (\ref{previous}) can be effectively described as dichotomic and treated as an additional qubit: $ \vert \bm{p} \rangle \equiv \vert 1 \rangle$ and $\vert  \bm{q} \rangle \equiv\vert 0 \rangle$. This simplified approximation shall be adopted in this section.

To describe transformation properties of quantum entanglement carried by pairs of physical particles in the context of a parity symmetric theory, such as the electron in QED, the Dirac bispinors must be used.  For example, the Bell-like spin state with momentum superposition constructed with vectors belonging to the irreps of the Lorentz group
\begin{equation}
\label{old}
\vert \Phi \rangle =  \big(\,\cos({\alpha})\, \vert \bm{p} \rangle_1 \, \otimes \vert \bm{q} \rangle_2 + \sin({\alpha})\, \vert \bm{q} \rangle_1 \, \otimes \vert \bm{p} \rangle_2\, \big) \otimes \big( \, \cos({\theta}) \, \vert+ \rangle \otimes \vert - \rangle + \sin({\theta})\vert - \rangle \otimes \vert + \rangle \, \big)
\end{equation}
has its correspondent on the bispinor level constructed as the superposition
\begin{eqnarray}
\label{statesuperpositionmomentum}
\Phi &=&\frac{1}{N} \Big[\, \cos({\alpha}) \big( \, \cos({\theta}) \, u(p, +) \otimes\big( u(q, -)  +  \sin({\theta}) \, u(p, -) \otimes u(q, +) \, \big)\otimes \vert \bm{p} \rangle_1\otimes \vert \bm{q} \rangle_2   \nonumber \\
&+& \sin({\alpha}) \big(\,  \cos({\theta}) \,  u(q, +) \otimes u(p, -) + \sin({\theta}) \, u(q, -) \otimes u(p, +) \, \big) \otimes \vert \bm{q} \rangle_1\otimes \vert \bm{p} \rangle_2\,\Big]
\end{eqnarray}
where $N$ is the normalization factor,  $\alpha$ is the superposition angle between the momenta and $\theta$ is the superposition angle between the spins. We emphasize that such state is a 6-qubit state as each particle carries 3-qubits: spin, intrinsic parity and (the discrete) momentum, and the description of its entanglement properties is even more involving than those presented for the state (\ref{stategeneralized}). We thus focus on recovering previous results presented on the literature and pointing out some striking differences arising from our treatment.

\begin{figure}[h]
\centering
\includegraphics[width=  17 cm]{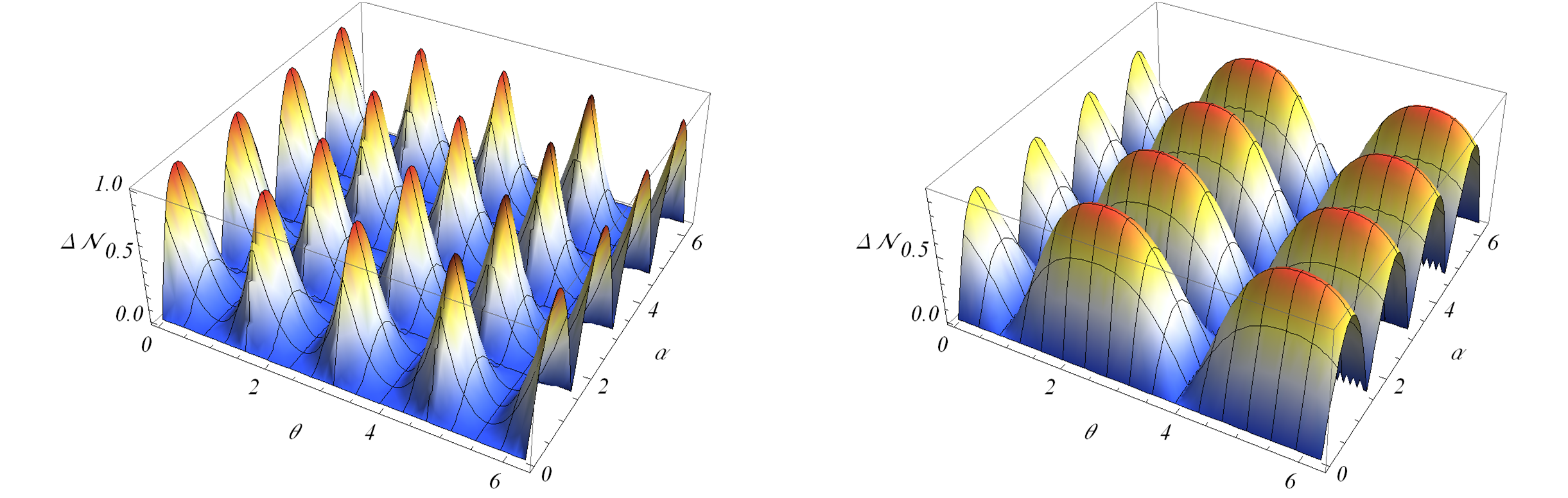}
\renewcommand{\baselinestretch}{1.0}
\caption{Entanglement in the partitions $\{S_1; \, p_1, \, S_2, \, p_2 \}$ (left plot) and $\{S_1, \, S_2; \, p_1, \, p_2 \}$ (right plot) for the positive parity projection (\ref{PP}) of the state (\ref{statesuperpositionmomentum}) under a Lorentz boost from the CoM perpendicular to the momenta. The initial and the boost rapidities are chosen to correspond to Wigner angles $\delta = \pi/2$ (left plot) and $\delta = \pi/4$. The variation of quantum entanglement in this partitions is exactly the same reported in the literature for spin-momentum entanglement encoded in Bell-like states (\ref{old}) belonging to the irreps of the Lorentz group under boosts \cite{relat07}.}
\label{Reproduction}
\end{figure}

The behavior of spin-momentum entanglement for (\ref{old}) under boosts can be reproduced by considering the projection of (\ref{statesuperpositionmomentum}) into positive parity states. The consideration of such projections discards the contribution of negative parity components of the bispinors, reducing the state vector to two non-vanishing components which, when the parity DoF is traced out, corresponds to the framework usually adopted the above mentioned transformation properties of entanglement. Starting from the full density matrix $\rho = \Phi \Phi^\dagger$, the projection into positive parity is obtained by
\begin{equation}
\label{PP}
\varrho = \frac{\hat{\Pi}_+^{(P)} \rho  \, \hat{\Pi}_+^{(P)}}{\mbox{Tr}[\hat{\Pi}_+^{(P)} \rho]},
\end{equation}
where $\hat{\Pi}_+^{(P)} = \vert + \rangle \langle + \vert _1 \otimes \vert + \rangle \langle + \vert _2 \otimes \hat{I}$ (with $\hat{I}$ the identity operator on the spins and momenta spaces). The entanglement between the spins and the momenta are evaluated through the reduced density matrix $\varrho_{S_1, p_1, S_2, p_2} = \mbox{Tr}_{P_1, P_2}[\varrho]$. For example, to compute entanglement between the spin of particle 1 and all other 3 DoFs, one compute the negativity $\mathcal{N}^{S_1; \, p_1, \, S_2, \, p_2}$ by the partial transposition of $\varrho_{S_1, p_1, S_2, p_2}$ with respect to $S_1$. In the same fashion, the entanglement between all spins and all momenta is evaluated by $\mathcal{N}^{S_1, \, S_2; \, p_1, \, p_2}$. The variation of $\mathcal{N}^{S_1; \, p_1, \, S_2, \, p_2}$ and $\mathcal{N}^{S_1, \, S_2; \, p_1, \, p_2}$ under boosts perpendicular to the momenta, from the CoM frame, is depicted in Fig.~\ref{Reproduction} in function of the superposition parameters $\alpha$ and $\theta$ and for parameters such that $\delta = \pi/2$ (left plot) and $\delta = \pi/4$ (right plot). The behavior of spin-momentum entanglement under boosts obtained with the above procedure is exactly the same reported in the literature \cite{relat07}, with a characteristic egg-tray behavior.

\begin{figure}[h]
\centering
\includegraphics[width=  9 cm]{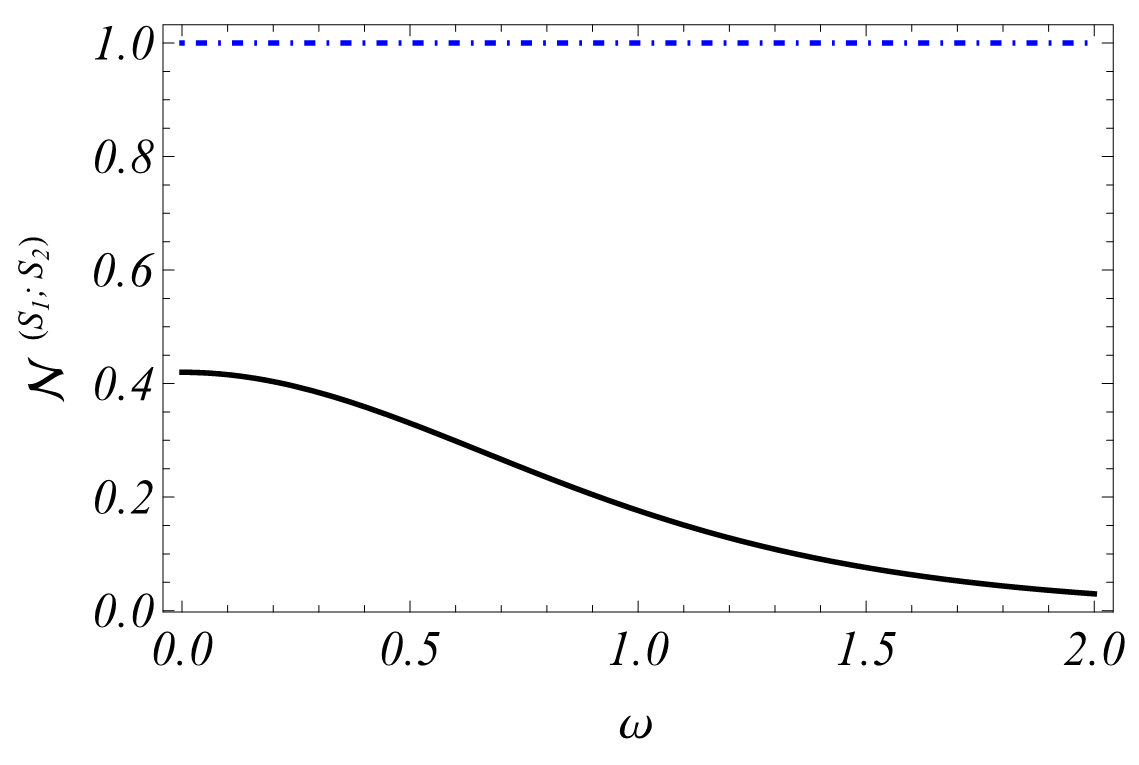}
\renewcommand{\baselinestretch}{1.0}
\caption{Variation of quantum entanglement between the spins of state (\ref{statesuperpositionmomentum}), for $\theta = \pi/4$ and $\alpha = \pi/4$, for a boost with rapidity $\omega$ perpendicular to the CoM frame. The solid line corresponds to the proper procedure to compute spin-spin entanglement, via the density matrix (\ref{trparity}) obtained by tracing the parity DoF. The dot-dashed line corresponds to the projection into positive parity (\ref{PP}) and corresponds to the framework usually considered in the literature. While by (\ref{trparity}) one predicts a degradation of the spin-spin entanglement, the consideration of only the posity parity component of the bispinorial state would lead to the conclusion that spin spin is invariant.}
\label{SS}
\end{figure}

Although the projection into positive parity states reproduces previous results, a correct characterization of spin-momentum entanglement in states such as (\ref{statesuperpositionmomentum}) is obtained by tracing out the parity DoFs, that is, by calculating the appropriate negativity of the density matrix
\begin{equation}
\label{trparity}
\rho_{S_1, p_1, S_2, p_2} = \mbox{Tr}_{P_1, P_2}[\rho].
\end{equation}
By construction, the state (\ref{statesuperpositionmomentum}) is momentum-spin separable, and thus  $\mathcal{N}^{S_1; \, p_1, \, S_2, \, p_2}[\rho_{S_1, p_1, S_2, p_2}] =\mathcal{N}^{S_1, \, S_2; \, p_1, \, p_2}[\rho_{S_1, p_1, S_2, p_2}] = 0$, but different from its counterpart (\ref{old}), a boost does not create entanglement in such partitions for any perpendicular boost from the CoM. This is a striking difference from the usual treatment that need to be addressed if one wants to describe entanglement encoded in the DoFs of a physical particle, such as an electron, a proton, a muon etc. A boost does not create spin-momentum entanglement in a separable Bell-like state and, if one wants to use correlations of such DoFs for some practical purpose, such as cryptography, such correlations need to be present from the beginning, for example, they must have been created by some scattering or decay process (see for example \cite{solano}). The significant difference between the results obtained by the procedures set by (\ref{PP}) and (\ref{trparity}) are also exhibited in other bipartitions of the system. For example, the entanglement encoded only between the spin DoFs, quantified by $\mathcal{N}^{S_1; \, S_2}$ depicted in Fig.~\ref{SS}, is invariant under boosts for (\ref{PP}), while for (\ref{trparity}) the boost generates a degradation in spin-spin entanglement similar to those reported in Figs.~\ref{Neg03},\ref{Neg01}.

\section{Conclusions and perspectives}

The behavior of entanglement encoded in spin states under Lorentz boosts has been the focus of recent research due to its importance in the implementation and characterization of physical protocols in setups involving reference frames relativistically moving with respect to each other \cite{relat01, relat02, relat03, relat04, relat05, relat06, relat07, relat08}. Usually, the quantum state considered is constructed with vectors belonging to the irreps of the Lorentz group, and the effects of Lorentz boosts are characterized by a rotation of the spin which depends on the momentum of the state \cite{wigner} -- a Wigner rotation. Although many interesting and insightful results were derived in this setup, when physical particles created by QED processes, such as electrons and protons, are considered, the states must be described in terms of the irreps of the complete Lorentz group \cite{wtung}, the Dirac bispinors.

This paper was concerned with the behavior of quantum entanglement encoded in superpositions of Dirac bispinors under Lorentz boost. The transformation law of bispinors is given by the exponential of the generator of the boost and it was supposed that, in the unboosted frame particles have momentum in the $e_z$ direction, and that the bispinors were helicity eigenstates. Due to the intrinsic spin-parity structure of Dirac bispinors \cite{SU2}, two particle states corresponds to four-qubit states, which can be entangled in several non equivalent ways. The entanglement between different bi-partitions of the systems was calculated in terms of the appropriate negativity, and as a global measures of entanglement  it was adopted the average negativity in each type of bi-partition. For example, the spin-spin entanglement, which can be used for quantum information purposes, was calculated through the negativity of the density matrix reduced to the spin-spin subsystem, obtained by tracing all other subsystems. In such multipartite set, the focus was on the study of the effects of Lorentz boosts on the different entanglements to investigate how this type of transformation redistribute the entanglement initially encoded in a set of bipartition among the other possible types, which can be directly related, for example, to the non-locality of the correlations encoded on the bispinors \cite{relat03, relat04} and as a characterization of the informational content of fermionic particles in a full relativistic setup \cite{bi-spinorarxiv02, celeri}. It is important to stress that the averages negativities provides a qualitative approach to the multipartite entanglement encoded among the different partitions of the systems once the quantification of multipartite entanglement measure is still an open problem. Two different boosts scenarios were considered: boosts parallel to the particle momenta, and boosts from the CoM frame perpendicular to the particle momenta. 

The first approach does not take into account momentum superposition. In this scenario each particle has a definite momentum and particle-particle entanglement, i.e. entanglement between all DoFs of one particle and all DoFs of the other particle, was shown to be invariant. On the other hand, other bi-partitions exhibit non-invariant behaviors. Spin-spin entanglement is degraded by the boost, being a non-monotonous function of the boost rapidity and exhibiting a maximum value at the CoM reference frame, in which entanglement between all intrinsic parities and all spins is minimal. As expected the entanglement between a given parity and all other DoFs vanish in the rest frame of one of the particles. For boosts perpendicular to the momenta, a similar behavior was observed for the negativities. The mean entanglement in bipartitions $\{i;\, j, \, k, \, l\}$ and $\{i; \, j, \, k\}$ increases under boosts, while $\{i, \, j; \, k, \, l\}$ and $\{i; \, j\}$ bi-partitions have a decreasing entanglement under Boosts. The bi-partition $\{i, \, j; \, k, \; l\}$ exhibit a non-monotonous behavior under boosts, decreasing for high-rapidity boosts.

Momentum superposition was also considered. Additionally to the superposition of the bispinors, it was supposed that the momenta of the particle were also superposed, and considering the simplified hypothesis of highly concentrated momenta distribution, allowing the effective description of the momentum DoF as an additional qubit, it was shown that, by considering the projection of the state into positive parity, i.e. by disregarding the negative parity components of the bispinors, it is possible to recover the egg-tray behavior of spin-momentum entanglement under boosts quoted in the literature \cite{relat07}. Although the consistence with this previous known results, the proper way to evaluate quantum entanglement in such scenario is by tracing out the parity DoF without any projection, and we showed that, in this case, the spin-momentum entanglement is an invariant quantity. Such striking difference between both methods is also present in the spin-spin entanglement which, for the correct method of tracing the parity DoFs, exhibit the same degradation observed previously, while for the projection method, this correlation is invariant. 

Considering that in the present literature quantum entanglement in relativistic scenarios is mostly described in the light of the irreps of the Poincar\'{e} group, our results set a new framework to discuss the transformation properties of quantum correlations in a complete covariant scenario, and are a fruitful and interesting addition to previous constructions that consider Dirac bispinors \cite{spins, bi-spinorarxiv02, bi-spinorFW}. Furthermore, given that bispinors naturally appear in QED processes, such as scatterings and creation/annihilation processes \cite{weinberg, shaw}, the framework developed in our paper can be further extended to the description of quantum entanglement in such context, which has also attracted recent attention \cite{scattering}.

{\em Acknowledgments - The work of AEB is supported by the Brazilian Agencies CAPES (grant 15/05903-4), FAPESP (grant 15/05903-4) and CNPq (grant 300809/2013-1). The work of VASVB is supported by the Brazilian Agency CAPES (grant 88881.132389/2016-1).}


\section*{References}

\begin{thebibliography}{99}

\bibitem{relat01}
Gingrich R M and Adami C 2002 Phys. Rev. Lett. {\bf 89} 270402. 

\bibitem{relat02}
Peres A, Scudo P F and Terno D R 2002 Phys. Rev. Lett. {\bf 88} 230402. 

\bibitem{relat03}
Ahn D, Lee H J, Moon Y H and Hwang S W 2003 Phys. Rev. A {\bf 67} 012103. 

\bibitem{relat04}
Terashima H and Ueda M 2003 Int. J. Quant. Inf. {\bf 1} 93. 

\bibitem{relat05}
Bartlett S D and Terno D R 2005 Phys. Rev. A {\bf 71} 012302. 

\bibitem{relat06}
Jordan T F, Shaji A and Sudarshan E C G 2007 Phys. Rev. A {\bf 75} 022101. 

\bibitem{relat07}
Friis N, Bertlmann R A, Huber M and Hiesmayr B C 2010 Phys. Rev. A {\bf 81} 042114. 

\bibitem{relat08}
Palge V and Dunningham J 2012 Phys. Rev. A {\bf 85} 042322. 

\bibitem{relatvedral}
Dunningham J, Palge V and Vedral V 2009 Phys. Rev. A {\bf 80} 044302; Palge V, Vedral V and Dunningham J. A. 2011 Phys. Rev. A {\bf 84} 044303. 


\bibitem{clock}
Jozsa R, Abrams D S, Dowling J P and Williams C P 2000 Phys. Rev. Lett. {\bf 85} 2010; Giovannetti V, Lloyd S and Maccone L 2001 Nature {\bf 412} 417; Yurtsever U and Dowling J P 2002 Phys. Rev. A {\bf 65} 052317. 

\bibitem{fonda}
Fonda L and Ghirardi G C 1970 \textit{Symmetry Principles in Quantum Physics} (New York: Marcel Dekker INC).

\bibitem{wtung}
Tung W K 2003 \textit{Group Theory} (London: World Scientific Publishing).

\bibitem{wigner}
Wigner E 1939 Ann. Math. {\bf 40} 149

\bibitem{weinberg}
Weinberg S 1995 \textit{Quantum theory of Fields vol. 1} (New York: Cambridge University Press).

\bibitem{spins}
Caban P, Rembielinski J and Wlodarczyk M 2013 Phys. Rev. A {\bf 88} 022119; Bauke H, Ahrens S, Keitel C H and Grobe R 2014 New Journal of Physics {\bf 16} 043012; Saldanha P L and Vedral V 2012 New Journal of Physics {\bf 14} 023041. 

\bibitem{bi-spinorarxiv02}
Alsing P M and Milburn G J [arXiv:quant-ph/020305]. 

\bibitem{bi-spinorFW}
Choi T, Hur J and Kim J 2011 Phys. Rev. A {\bf 84} 012334; Choi T 2013 Journal of the Korean Physical Society {\bf 62} 1085. 

\bibitem{SU2}
Mizrahi S S 2009 Physica Scripta {\bf T135} 014007; Bernardini A E and Mizrahi S S 2014 Physica Scripta {\bf 89} 075105.

\bibitem{greiner}
Greiner W 2000 \textit{Relativistic Quantum Mechanics: Wave equations} (Berlin: Springer).

\bibitem{reventanglement}
Horodecki R, Horodecki P, Horodecki M. and Horodecki K. 2009 Rev. Mod. Phys. {\bf 81} 865.

\bibitem{neg01}
Peres A 1996 Phys. Rev. Lett. {\bf 77} 1413.

\bibitem{neg02}
Vidal G and Werner R F 2002 Phys. Rev. A {\bf 65} 032314.

\bibitem{aop01}
Bittencourt V A S V, Mizrahi S S and Bernardini A E 2015 Annals of Physics {\bf 355} 35-47.

\bibitem{aop02}
Bittencourt V A S V and Bernardini A E 2016 Annals of Physics {\bf 364} 182. 

\bibitem{diraclike01}
Bittencourt V A S V, Bernardini A E and Blasone M 2016 Phys. Rev. A {\bf 93} 053823. 

\bibitem{diraclike02}
Bittencourt V A S V and Bernardini A E 2017 Phys. Rev. B {\bf 95} 195145.

\bibitem{rigolin}
Rigolin G, de Oliveira T R and de Oliveira M C 2006 Phys. Rev. A {\bf 74} 022314.

\bibitem{MassimoEnt}
Blasone M, Dell'Anno F, De Siena S and Illuminati F 2008 Phys. Rev. A {\bf 77}, 062304. 

\bibitem{celeri}
C\'{e}leri L C, Kiosses V and Terno D R 2016 Phys. Rev. A {\bf 94} 062115. 

\bibitem{NewPRA}
Bittencourt V A S V, Bernardini A E and Blasone M 2018 arXiv:1801.00758[quant-ph].

\bibitem{solano}
Pachos J and Solano E 2003 Quant. Info. Comm. {\bf 3} 115. 

\bibitem{shaw}
Mandl F and Shaw G 2010 \textit{Quantum Field Theory} (West Sussex: John Wiley \& Sons)

\bibitem{scattering}
Peschanski R and Seki S 2016 Phys. Lett. B {\bf 758} 89; Ratzel D, Wilkens M and Menzel R 2017 Phys. Rev. A {\bf 95} 012101; Fan J, Deng Y and Huang Y-C 2017 Phys. Rev. D {\bf 95} 065017.





\end{thebibliography}
\end{document}